**Neural measures of the causal role of observers' facial mimicry on visual working memory for facial expressions**

**Paola Sessa[1,2*], Arianna Schiano Lomoriello[1] and Roy Luria[3,4]**

[1] Department of Developmental and Social Psychology, University of Padova, Padova, Italy

[2] Padova Neuroscience Center (PNC), University of Padova, Padova, Italy

[3] School of Psychological Science, Tel Aviv University, Tel Aviv 6997801, Israel

[4] Sagol School of Neuroscience, Tel Aviv University, Tel Aviv 6997801, Israel

**\* Corresponding author:**

Paola Sessa

Associate Professor

Department of Developmental and Social Psychology (DPSS) and Padova Neuroscience Center (PNC), University of Padova, Via Venezia 8, 35131 Padova (Italy)

Phone: (+39) 0498277455, Fax: (+39) 0498276511

Cognition and Language Research Laboratory (CoLab): http://colab.psy.unipd.it

Padova Neuroscience Center (PNC): http://pnc.unipd.it/

**Key words:** facial mimicry, facial expressions, face processing, visual working memory, SPCN, CDA, empathy, event-related potentials, simulation

**Abstract**

Simulation models of facial expressions propose that sensorimotor regions may increase the clarity of facial expressions representations in extrastriate areas. We monitored the event-related potential marker of visual working memory (VWM) representations, namely the sustained posterior contralateral negativity (SPCN), also termed contralateral delay activity (CDA), while participants performed a change detection task including to-be-memorized faces with different intensities of anger. In one condition participants could freely use their facial mimicry during the encoding/VWM maintenance of the faces, while in a different condition, participants had their facial mimicry blocked by a gel. Notably, SPCN amplitude was reduced for faces in the blocked mimicry condition when compared to the free mimicry condition. This modulation interacted with the empathy levels of participants such that only participants with medium-high empathy scores showed such reduction of the SPCN amplitude when their mimicry was blocked. The SPCN amplitude was larger for full expressions when compared to neutral and subtle expressions, while subtle expressions elicited lower SPCN amplitudes than neutral faces. These findings provide evidence of a functional link between mimicry and VWM for faces, and further shed light on how this memory system may receive feedbacks from sensorimotor regions during the processing of facial expressions.

**Introduction**

Humans are incredibly efficient in understanding the affective states of others and in empathizing with them, and in particular they are exceptionally capable of inferring others' affective state by extracting this information from a single look at their face. Among the stimuli able to convey information on the emotions of others, faces indeed occupy a place of prime importance and for this reason faces in general, and facial expressions in particular, are of paromount importance for individuals. According to recent models aimed at explaining this process, the understanding of others' emotions expressed through their faces would be carried out through a simulation of those emotions by the observer (Carr et al., 2003; Caruana and Borghi, 2013; Gallese and Sinigaglia, 2011; Goldman and and de Vignemont, 2009; Goldman and Sripada, 2005; Niedenthal, 2007; Pitcher, Garrido et al., 2008; Wicker et al., 2003; see also Mastrella and Sessa, 2017, for a review on this topic). This simulation process would also involve the facial mimicry of the observer, and depending on the specific simulation model, this mimicry would have a different role in order to extract the meaning of the emotion from a visual pattern linked to a facial expression. Some models assign to mimicry an absolutely central role in this process, others indicate it as a fundamentally accessory element (see Goldman and Sripada, 2005 for a review of different simulation models; see also Hess and Fischer, 2014 for a review on emotional mimicry) or a sort of spillover of simulation by sensorimotor areas (Wood et al., 2016). In favor of the link between mimicry and simulative processes, emotional mimicry has been associated with the activation of the mirror neuron system such that congruent facial reactions to angry and happy expressions correlate significantly with activations in the inferior frontal gyrus, supplementary motor area and cerebellum (see Likowski et al., 2012).

To note, the concept of mimicry has also been closely linked to emotional contagion in particular, and to the construct of empathy more specifically (see Prochazkova and Kret, 2017, for a recent review on this topic). Empathy is a multifaceted construct that consists of at least two different aspects: a "neural resonance" process, which is the automatic tendency to simulate the states of others, including sensory, motor and emotional components, and a mentalizing process, through which would be possible to explicitly assign affective states to others (see, e.g., Bruneau et al., 2015; Decety and Lamm, 2006; Decety and Svetlova, 2012; Kanske et al., 2016; Kanske et al., 2015; Lamm et al., 2011; Marsh, 2018; Meconi et al., 2018; Meconi et al., 2015; Sessa et al., 2014; Vaes et al., 2016; Zaki and Ochsner, 2012). For instance, in a study by Sonnby-Borgström (2002) the participants were exposed to angry, neutral and happy facial expressions, while their facial movements and reactions were recorded using electromyography (EMG). The authors aimed at investigating how facial mimicry behavior in "face-to-face interaction" situations was related to individual differences in emotional empathy. The results of this study demonstrated that at individual differences in empathy correspond differences in the zygomaticus muscle reactions. Specifically, the high-empathy group was characterized by a significantly higher correspondence between facial expressions and self-reported feelings; on the contrary, the low-empathy group showed inverse zygomaticus muscle reactions, namely "smiling" when exposed to an angry face. Interestingly, no differences were found between the high- and low-empathy subjects in their verbally reported feelings when presented a happy or an angry face. Thus, the differences between the groups in emotional empathy appeared to be related to differences in automatic somatic reactions to facial stimuli rather than to differences in their conscious interpretation of the emotional situation.

The objective of the present investigation was to test whether the mimicry of the observer is a critical element for the construction of visual working memory (VWM) representations of facial expressions of emotions, monitoring whether this process can also depend on the degree of empathy of the observer. The VWM buffer constitutes a critical hub between these earlier processing stages and the manifest behavior of the individuals (see, e.g., Luck, 2005). Thus, demonstrating that the observer's facial mimicry could have an impact on the functioning of this buffer is of fundamental importance for understanding how facial expressions recognition occurs, and in particular how the VWM buffer operates when coordinating online behavior.

The concertation of processing stages/activity in different brain regions necessary for an understanding of others' emotions is well delineated by a recent theoretical model proposed by Wood and colleagues (2016) which considers the recognition/discrimination of others' emotions as a complex process involving the parallel activation of two different systems, one for the visual analysis of faces and facial expressions, and a second one for sensorimotor simulation of facial expressions. This second system would trigger the activation of the emotion system, that is the whole of those additional brain regions involved in the emotional processing, including the limbic areas. The combination of these processing steps, in continuous iterative interaction with each other, would allow us to understand the emotion expressed by others' faces, assigning an affective state to the others, and possibly producing appropriate behavioral responses. Within this theoretical framework, a large body of studies supports a central role of the observer's facial mimicry during emotion recognition and discrimination, such that, for instance, when mimicry is blocked/altered either mechanically (Baumeister et al., 2015; Niedenthal et al., 2001; Oberman et al., 2007; Stel and van Knippenberg, 2008; Wood et al., 2015) or chemically (for example by the botulinum toxin A-BTX; Baumeister et al., 2016), or in patients with facial paralysis (Keillor et al., 2002; Korb et al., 2016) emotional faces recognition is disturbed.

A crucial aspect of Wood and colleagues' model (2016) is that the sensorimotor simulation process (which, according to the authors, may or may not involve the facial mimicry of the observer depending on the intensity of the simulation) "feeds back to shape the visual percept itself" (Wood et al., 2016). This aspect of the model therefore implies that interfering with the simulation mechanism may have an effect on the quality of the representation of facial expressions. However, at the moment there is no direct evidence in literature of this specific fascinating aspect of Wood and colleagues' model.

The present experimental investigation aimed at testing the aspect of the model by Wood and colleagues (2016) that hypothesized a feedback process from simulation to facial percept representation. In particular, the main research question that led the present study was whether alteration/blocking of facial mimicry by using a hardening gel mask could interfere with VWM representations of emotional facial expressions. This would allow first of all to demonstrate the feedback processing postulated by the model, in which the simulation has an effect on the construction of facial expression representations, and also would allow a better understanding of the functioning of the VWM buffer, since an effect of block/alteration of mimicry on this buffer would demonstrate that the simulation process normally contributes to its functioning in the case faces are represented/stored.

For this purpose, we implemented a variant of the classic change detection task (Luria et al., 2010; Meconi et al., 2014; Sessa and Dalmaso, 2016; Sessa et al., 2011; Sessa et al., 2012; Vogel and Machizawa, 2004; Vogel et al., 2005) in which participants, who wore the gel mask for half of the experiment (manipulation of the gel within-subjects), were asked, in each trial, to memorize a face (*memory array*) with a facial expression of three possible intensities (neutral, subtle, intense) for a short time interval of about 1 sec, and to decide, at the presentation of a *test array*, if the expression of the presented face was the same or different from that of the memorized face. The identity of the faces did not change within the same trial. In order to monitor the event-related potential (ERP) named Sustained Posterior Contralateral Negativity (SPCN; Jolicœur et al., 2007; Luria et al., 2010; Meconi et al., 2014; Sessa and Dalmaso, 2016; Sessa et al., 2011, 2012) or also Contralateral Delay Activity (CDA; Vogel and Machizawa, 2004) the presentation of the stimuli was lateralized in the visual field and a distractor face was presented on the opposite side. An arrow placed in the center of the screen, immediately above the fixation cross, indicated to the participants – for each trial – if they had to memorize the face in left or right side of the memory array. The participants then had to compare this memorized face with the face that appeared in the same position in the test array.

The SPCN/CDA is a well-known marker of VWM representations (see Luria, Balaban, Awh,

and Vogel, 2016 for a review). It is defined as the difference between the activity recorded at posterior sites (PO7/PO8, O1/O2, P3/P4) contralaterally to the presentation of one or more to-be-memorized stimuli and the activity recorded ipsilaterally to the presentation of one or more to-be-memorized stimuli. SPCN/CDA amplitude tends to increase as the amount of information to be memorized or the quality (in terms of resolution) of the representation increases (e.g., Sessa et al.,

2011, 2012; Vogel and Machizawa, 2004) until an asymptotic limit is reached that corresponds to the saturation of the VWM capacity (e.g., Vogel and Machizawa, 2004).

In one of our previous studies (Sessa et al., 2011) we have shown that faces with intense

facial expressions (fear) elicit larger SPCN/CDA than faces with the same identity but with neutral

expressions. These findings suggest either that emotions are represented in VWM, such that VWM

representations may include the emotion information, or that faces with an emotional expression are represented with higher resolution than neutral faces (see also Stout, Shackman, and Larson, 2013, for a replication of these findings).

The SPCN proves therefore to be a very useful marker in the present experimental context,

since it offers us the opportunity to test the role of obervers' facial mimicry on the construction of facial expression representations in VWM. As a secondary goal, we wanted also to provide an extension of our previous results to a different negative facial expression (i.e., anger).

We expected to observe lower SPCN/CDA amplitude values for facial expressions

memorized when participants wore the hardening gel compared to the condition in which

participants' facial mimicry was not blocked/altered, that is, as suggested by Wood and colleagues'

model (2016) we hypothesized that mimicry may increase the clarity of visual representations of facial expressions, and as a consequence we expected higher resolution for facial expressions memorized when participants' mimicry could be freely used than when blocked/altered. We also expected larger SPCN/CDA amplitude values for intense expressions of anger when compared to both neutral and moderate expressions of anger (i.e., higher resolution for intense angry expressions than for neutral and subtle expressions).

Finally, as mentioned above, the literature strongly suggests that mimicry is stricly linked to empathy (see Prochazkova and Kret, 2017). On the basis of the evidence provided by the literature on the link between mimicry and empathy, we included the empathy variable in our experimental design by measuring participants' Empathic Quotient (Baron-Cohen and Wheelwright, 2004) in order to evaluate whether the interference on the simulation process by blocking/altering participants' mimicry may affect VWM representations of faces differently in high and low empathic individuals.

**Method**

Participants

Before starting with data collection, we decided to proceed with the analysis of a sample of about 30 participants, as the existing literature in the field suggests being an appropriate sample (refs).

The analyses were conducted only after completing the data collection. Data were collected from 36 healthy volunteer students of the University of Padova. Due to an excess of electrophysiological artifacts, especially eye movements, data from 7 participants were discarded from the analysis. All participants reported normal or correct vision from lenses and no history of neurological disorders. Twenty-nine participants (18 males, average age in years = 24, SD = 2.73; 2 left handed) were included in the final sample.

Stimuli

The stimuli were grayscale digital photographs of faces of 8 individuals (4 females and 4 males) expressing three different levels of emotional facial expression (neutral, sublte and full). These face stimuli have been modified by Vaidya, Jin and Fellows (2014) from the original images taken from the Karolinska database (Lundqvist et al., 1998) that included neutral and angry facial expressions. The subtle facial expressions were generated by a morphing procedure of facial expressions of the neutral and angry expressions of the same individual from Vaidya, Jin and Fellows (2014) and were 30-40% along the morph continuum. Figure 1 shows the three levels of facial expressions' intensity for 2 individual faces (one of a female and one of a male). All images have been resized to subtend at a visual angle between 10 and 12 degrees. The participants were seated at 70 cm from the screen. The stimuli were presented on a 7" inch CRT monitor of a computer with E-prime software.

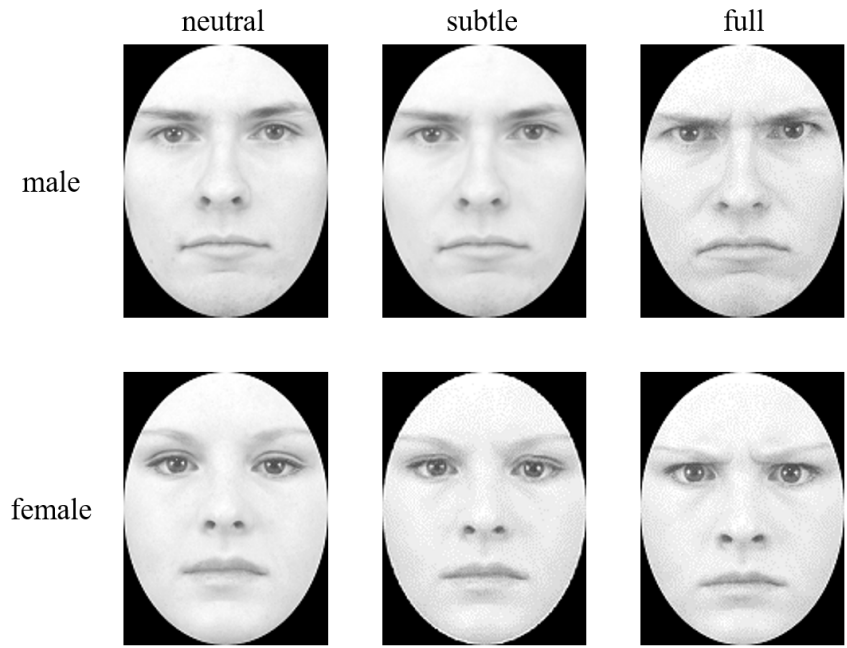

Procedure

We used a variant of the change detection task (e.g., Sessa et al., 2011; Vogel and Machizawa, 2004). Each trial began with a fixation of 500 ms, that remained in the center of the screen throughout the trial, followed by the presentation of two arrows as cues shown for 200 ms one above and one below the fixation cross, both pointing in the same direction (ie, both on the left or both on the right). The two cues, or the two arrows, were shown for 200 ms and followed by a blank screen of variable duration (200–400ms). Then a memory array of faces appeared, presented for 500 ms. The memory array consisted of two faces with a neutral, subtle or full emotional facial expression of anger.

Following the memory array, a blank screen with a duration of 900 ms preceded the test array onset, which also contained two faces, one on the right and one on the left of the fixation cross, which was shown until an answer was provided by the participant. In the memory and in the test array, faces of the same identity were presented. Participants were instructed to maintain their gaze on the fixation cross throughout the trial and to memorize only the face of the memory array shown in the side indicated by the arrows and were also explicitly informed that the face shown on the opposite side was not relevant for the task at hand. The task was to compare the memorized face with the one presented on the same side of the test array, in order to indicate if the facial expression of the face had changed or not. In 50% of the trials, the facial expression in the memory array and the test array were identical. In the remaining 50% of the trials, the facial expression was replaced in the test array with a different facial expression. When a change occurred, the face stimulus was replaced with a face stimulus of the same individual but which presented a different intensity of facial expression.

Half of the participants were instructed to press the "F" key to indicate a change between the memory array and the test array and the "J" key to indicate that there was no change between the memory array and the test array. The other half of the participants responded on the basis of an inverted response mapping. The responses had to be given without any time pressure: the participants were informed in this regard that the speed of response would not be taken into account for the evaluation of their performance. Following the response, a variable interval of 1000−1500 ms (in 100 ms steps) elapsed before the presentation of the fixation cross indicating the beginning of the next trial. The experiment started with a block of 12 trial trials. The participants performed 4 experimental blocks, each of 144 trials (i.e., 432 trials in total). Figure 2 shows the trial structure of the change detection task.

Each participant has performed the task in two different conditions (counterbalanced order across the participants); in the gel condition a mask gel was applied on the participant's whole face, so as to create a thick and uniform layer, excluding the areas near the eyes and upper lip. The product used as a gel was a removable cosmetic mask (BlackMask Gobbini©) that dries in 10 minutes from application and becomes a sort of plastified and rigid mask. The participants perceived that the gel prevented the wider movements of face muscles. In the other half of the experiment (no-gel condition) nothing was applied to the participants' faces.

As in the study by Wood et al. (2015), at the beginning of the experimental session participants were told that the experiment involved "the role of skin conductance in perception" and that they would be asked to spread a gel on their face in order to "block skin conductance" before completing a computer task.

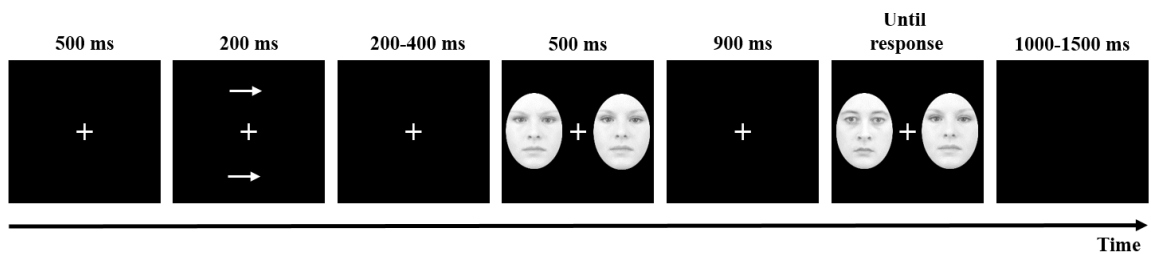

### EEG/ERP recording

The EEG was recorded during the task by means of 64 active electrodes distributed on the scalp according to the extended 10/20 system, positioning an elastic Acti-Cap with reference to the left ear lobe. The high viscosity of the gel used has allowed the impedance to be kept below 10 KΩ. The EEG was re-referenced offline to the mean activity recorded at the left and right ear lobes. The EEG has been segmented into epochs lasting 1600 ms (-200/1400). Following the baseline correction, trials contaminated by ocular artifacts (i.e. those in which the participants blinked or moved the eyes, eliciting activity higher than ± 30 µV or ± 60 µV, respectively) or from other types of artifacts (greater than ± 80 µV) were removed. Finally, the contralateral waveforms were computed by mediating the activity recorded by the electrodes of the right hemisphere when the participants were required to encode and memorize the face stimulus presented on the left side of the memory array (pooling of the electrodes O2, PO8, P4 ) with the activity recorded by the electrodes positioned on the left when the participants were required to encode and memorize the face stimulus presented on the right side of the memory array (pooling of electrodes O1, PO7, P3). The SPCN was quantified as the difference in mean amplitude between the contralateral and ipsilateral waveforms in a time window of 300-1300 ms time-locked to the presentation of the memory array for each experimental condition (facial expression: neutral, subtle, full; condition: gel and no-gel).

At the end of the EEG session involving the change detection task, the participants were given the Empathy Quotient questionnaire (Baron-Cohen and Wheelwrigh, 2004). The EQ measures the empathic skills of the individual through 80 items (20 of which are control items). Individuals have to express their agreement on a 4-point Likert scale: "very much agree", "partially agree", "partially disagree" and "very much in disagreement". The analysis of the scores is carried out on the basis of a scale ranging from 0 (almost null empathy) to 80 (exceptional empathy).

The EQ values were then sorted in ascending order and the participants were divided into 2 groups so that a group of participants had a medium-low EQ average value ($N = 15$) and another group a medium-high EQ average value ($N = 14$). The rationale for this procedure was based on the assumption that individuals with higher empathic abilities are more likely to use their facial mimicry when recognizing and discriminating other people's facial expressions than individuals with lower empathic abilities (e.g., (Sonnby-Borgström, 2002; Sonnby-Borgström, Jönsson, and Svensson, 2003). From this point of view it is possible that the blocking/altering facial mimicry by means of the gel could compromise the representations of facial expressions more in the participants with medium-high EQ than in the participants with medium-low EQ.

### Results

Behavior

The mean proportion of correct responses was submitted to an analysis of variance (ANOVA) considering the within-subject factors emotion (neutral, subtle, full), the mimicry condition (free vs. blocked/altered by the presence of the gel) the between-subjects factor EQ (medium-low EQ, medium-high EQ). The only statistically significant effect was that of emotion, $F(2,26) = 325.937$, $p < .001$, $\eta p^2 = .923$. Following planned comparisons indicated that participants were more accurate when they had to memorize faces with full expressions than when they had to memorize faces with neutral expressions ($p < .001$, $SE = .008$, 95% $CI$ [.123, .163]) or subtle expressions ($p < .001$, $SE = .007$, 95% $CI$ [.160, .196]). In addition, the participants were more accurate when they had to memorize faces with neutral expressions than faces with subtle expression ($p < .001$, $SE = .007$, 95% $CI$ [.016, .054]). The effect of the mimicry condition and the interaction between emotion and mimicry conditions were not statistically significant ($F = 1.463$, and $F < 1$, respectively). See the Figure 3.

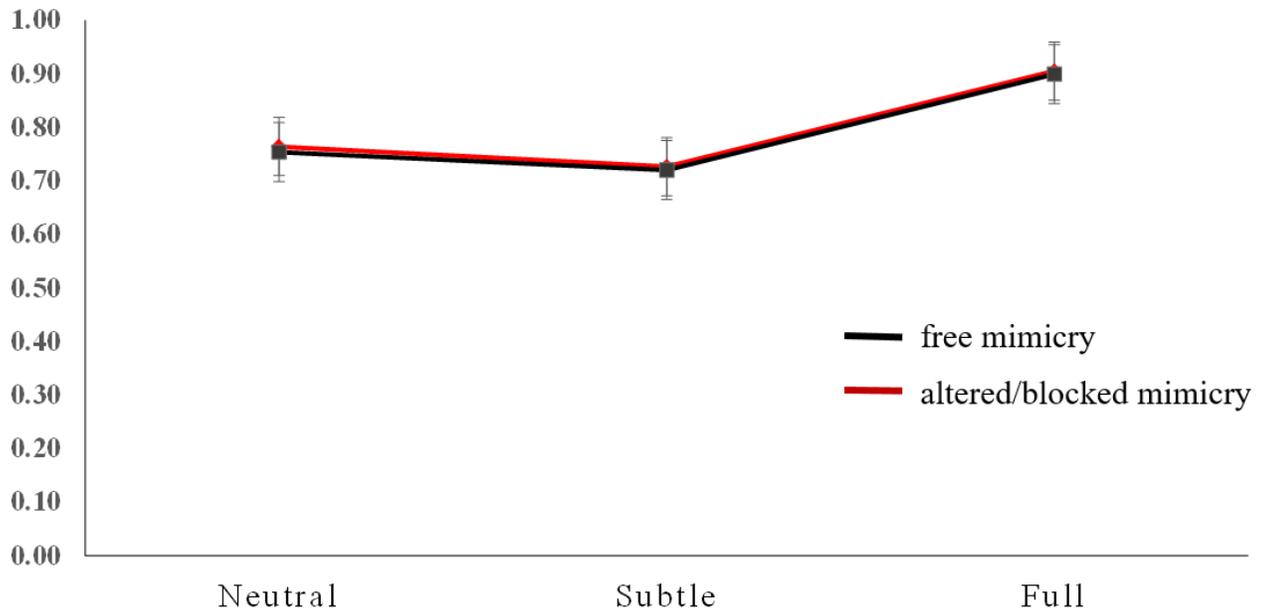

SPCN

An ANOVA of the mean SPCN amplitude values was performed including the within-subjects factors emotion (neutral, subtle, full) and mimicry condition (free vs. blocked/altered by the presence of the gel) and the between-subjects factor EQ (medium-low EQ, medium-high EQ). Wherever appropriate the Greenhouse-Geisser correction was used.

The ANOVA revealed a significant main effect of the emotion ($F(2,26) = 9.841$, $p = .001$, $\eta p^2 = .267$), of the mimicry condition ($F(1,27) = 5.189$, $p =. 031$, $\eta p^2 = .161$)), and an interaction between the mimicry condition and the EQ ($F(1,32) = 4.617$, $p = .041$, $\eta p^2 = .146$). The other interactions were not statistically significant ($F < 1$). Pairwise comparisons revealed that facial expressions of full anger elicited larger SPCN amplitude values (mean SPCN amplitude for full expressions = -1.03 µV) when compared to both neutral expressions ($p = .004$, $SE = .104$, 95% $CI$ [-.540, -.112] mean SPCN amplitude for neutral expressions = -.71 µV) and subtle expressions ($p = .001$, $SE = .166$, 95% $CI$ [.956, -.275]; mean SPCN amplitude for subtle expressions = -.42 µV). Interestingly, subtle expressions produced reduced SPCN amplitude values when compared to neutral expressions ($p$ = .047, $SE$ = .139, 95% $CI$ [.004, .575]). In brief, the whole pattern of SPCN mean amplitudes elicited by the different levels of emotions nicely mirrored participants' accuracy in the change detection task. The differential waveforms (contralateral-minus-ipsilateral) time-locked to the presentation of the memory array for each level of facial expression (neutral, subtle, full) are presented in Figure 4.

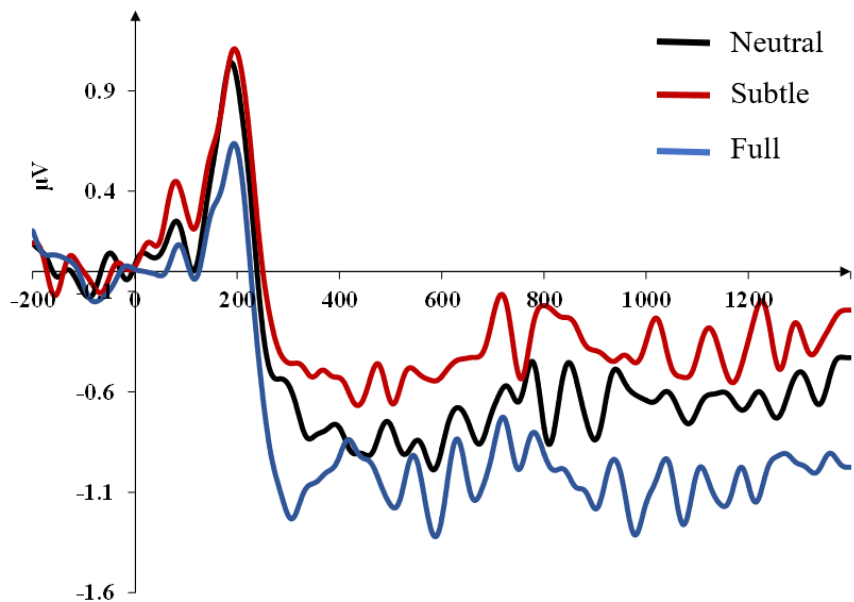

Pairwise comparisons (Bonferroni corrected) for participants with medium-low EQ did not highlight any effect of the mimicry condition ($F$ < 1, $SE$ = .183, 95% $CI$ [-.391, .358]; mean SPCN amplitude for the blocked/altered mimicry condition = -.73 μV, for the free mimicry condition = -.75 μV), but, importantly, participants with medium-high EQ showed that the blocked/altered mimicry significantly impacted the SPCN amplitude ($F$ = 9.471, $p$ = .005; $SE$ = .189, 95% $CI$ [-.969, .194]; $\eta p^2$ = .260; mean SPCN amplitude for the blocked/altered mimicry condition = -.40 μV, for the free mimicry condition = -.98 μV).

Figure 5 shows the differential waveforms (contralateral-minus-ipsilateral) time-locked to the presentation of the memory array for the two mimicry conditions (free vs. blocked/altered) for medium-low EQ participants (panel A) and medium-high EQ participants (panel B) separately.

These results therefore suggest that VWM representations of facial expressions appear to be impaired by the gel in the more empathic participants, but not in the less empathic participants.

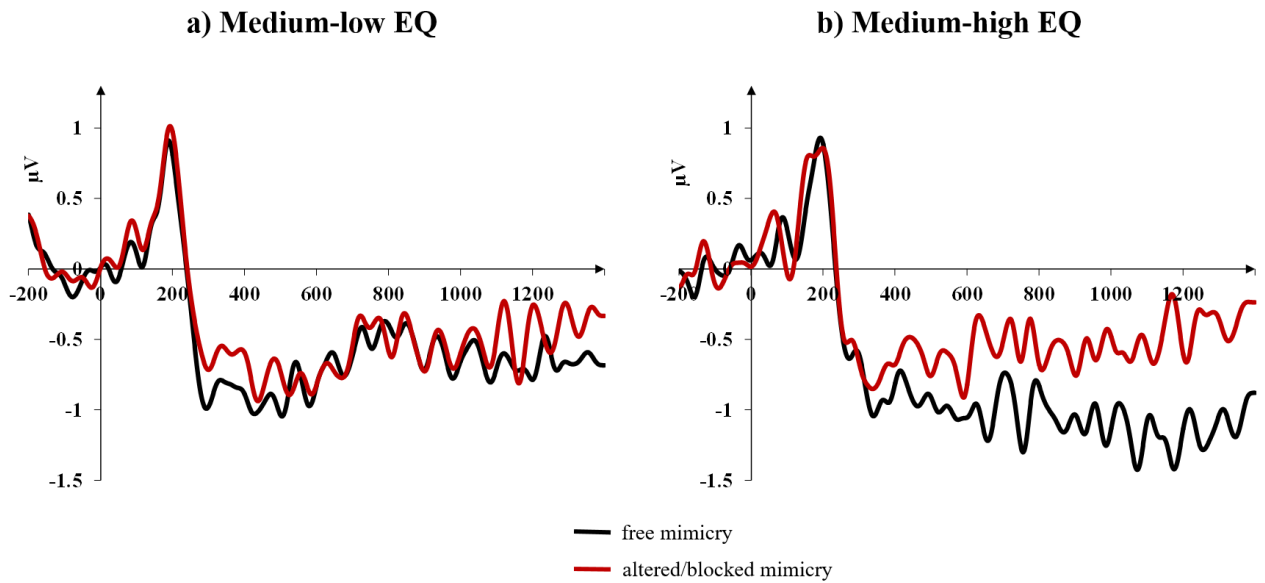

**a) Medium-low EQ**    **b) Medium-high EQ**

—— free mimicry
—— altered/blocked mimicry

## Discussion

The present experimental investigation, based on the theoretical background offered by the simulation models of facial expressions and, more specifically, on the model recently proposed by Wood and colleagues (2016), had the objective of testing whether the facial mimicry of an observer could be an element able to modulate the visual representations of facial expressions, as predicted by this latter model that proposes a feedback processing from simulation in sensorimotor areas to processing in the extrastriate areas, such that the simulation process is able to increase the clarity of visual representations of facial expressions.

With the aim to test this hypothesis, we implemented a variant of the classic change detection task in order to monitor the electrophysiological marker of VWM representations, namely the SPCN/CDA ERP component, in two critical experimental conditions in a within-subjects design. In one experimental condition the participants performed the change detection task, that included faces with different intensities of facial expression of anger (neutral, subtle and full) while being able to freely use their facial mimicry during the encoding and VWM maintenance of the face stimuli; in a different experimental condition, critical to test the main hypothesis of the present investigation, the participants performed the same task but their facial mimicry was blocked/altered by a facial gel that, hardening, greatly limited their facial movements.

In line with our hypotheses, the results with regard to the gel manipulation showed reduced SPCN/CDA amplitude values for the face representations stored when participants wore the facial gel compared to when their mimicry was not blocked/altered, thus suggesting that the information maintained in VWM under the blocked/altered mimicry condition was poorer than that maintained in the condition in which participants could naturally use their facial muscles during the exposure to facial expressions.

We did not observe differences between the two conditions of anger expressions (subtle vs. full) in terms of modulation of the SPCN amplitude. One possible explanation is that mimicry is not so selectively sensitive to different levels of negative expressions, as for instance it is suggested by a recent study by Fujimura, Sato, and Suzuki (2010). These authors have indeed provided experimental evidence that the intensity level (arousal) of facial expressions induces in the observers a modulation of their facial mimicry (measured as the electromyography reactions of the zygomatic major and supercilii corrugator muscles) only in the case of expressions with positive valence, but not in the case of negative expressions that instead induced the same level of mimicry in the observers, as indicated by the activity of their supercilii corrugator muscle. On the other hand, another possibility that we cannot disregard is that the statistical power of our study could have been able to allow observing an overall effect of the mimicry but not more subtle modulations related to the different levels of facial expressions.

A second main objective of our work was also to investigate whether the level of empathy of the observer could be an important variable for understanding the role of mimicry on the construction and maintenance of facial expression representations in VWM. The evidence in the literature that guided this hypothesis strongly suggests that individuals with higher levels of empathy are more likely to use their facial mimicry during exposure to facial expressions such as happiness and anger (Dimberg, Andréasson, and Thunberg, 2011; Sonnby-Borgström, 2002; Sonnby-Borgström et al., 2003), and activate their corrugator muscle even when exposed to fear (Balconi and Canavesio, 2016) and disgust (Balconi and Canavesio, 2013; see also Rymarczyk, Zurawski, Jankowiak-Siuda, and Szatkowska, 2018). These findings, together, constitute an important body of knowledge that supports the view that mimicry is an important component of emotional empathy (see, e.g., Preston and de Waal, 2002; Prochazkova and Kret, 2017). Our results are very nicely in agreement with this previous evidence; in fact the participants who suffered the most impairment of VWM facial expression representations (in terms of SPCN amplitude) due to the block/alteration of their facial mimicry were those with higher levels of empathy (as measured by the EQ).

A very recent study by de la Rosa, Fademrecht, Bülthoff, Giese, and Curio (2018) has used a clever paradigm of visual and motor adaptations to explore their effects on recognition of facial expressions. Notably, their results showed that visual adaptation (through the repeated visual presentation of facial expressions) and motor adaptation (through the repeated execution of facial expressions) had an opposite effect on expression recognition. These findings support a dual system for the recognition of others' facial expressions, one 'purely' visual and one based on simulation. Our results, nevertheless, seem to suggest that this dissociability does not imply that the two systems cannot influence each other, at least as regards the effect of the simulation system on the visual system. Furthermore, they suggest that there might be an important inter-individual variability in the connectivity of the two systems, such that this influence of the simulation system on the visual system (in our present work in terms of VWM representations) is particularly relevant for those individuals with higher levels of empathy who likely tend to recognize others' emotional expressions through the synergy of the two systems. Moreover, in the light of these observations, it should probably be emphasized that these findings inform us about the recognition/discrimination of emotions under conditions of interference with simulation, but nothing tells us about the observer's subjective experience that might differ in cases in which facial expressions processing is accomplished on a visual basis, on a simulation basis or through an integration of the two systems. We believe that answering this question is one of the most ambitious challenges that future research will have to face.

The present study also allowed us to replicate our previous findings (Sessa et al., 2011) regarding the effect of facial expressions on VWM representations extending those previous findings to a different negative facial expression, i.e. anger. In our previous work we have demonstrated that faces with an intense expression of fear elicit larger SPCN amplitudes than faces with a neutral expression, suggesting that fearful faces are either represented in greater detail (i.e., high-resolution representations) or that the emotion is also represented in VWM in some kind of additional visual format. In the present study, we have replicated these results with angry faces compared to neutral faces: we observed that facial expressions of intense anger elicited an SPCN of greater amplitude than neutral and subtle facial expressions and, moreover were associated with greater accuracy in the change detection task when compared to the other two levels of intensity of facial expressions (neutral and subtle). This overall pattern of findings is very well in line with the benefit observed for negative and angry facial expressions in previous studies in terms of behavioral indices of sensitivity (Jackson, Linden, and Raymond, 2014; Jackson, Wolf, Johnston, Raymond, and Linden, 2008; Langeslag, Morgan, Jackson, Linden, and Van Strien, 2009; Simione et al., 2014; Xie and Shang, 2016). An unexpected result refers to the observation that the faces with expression of subtle anger elicited not only an SPCN of lower amplitude than the faces with full expressions of anger, but also an SPCN of lower amplitude than the faces with neutral expressions. This electrophysiological pattern also entirely parallels accuracy in the change detection task, such that accuracy associated with trials subtle faces expressions was lower than accuracy for faces with both full and neutral expressions. A possible account for this result could take into consideration the concept of *distinctiveness*, originally coined in the context of long-term memory studies (see, e.g., Eysenck, 1979) and later considered a variable of great importance also in the context of short-term memory studies (Hunt, 2006). Distinctiveness refers to the ability of an item to produce a reliable representation in memory relative to the other items in a certain (experimental) context. In relation to the stage of information retrieval, distinctiveness specifies that it is more likely to recover memories/representations that are sparsely represented within the space of representation than memories/representations that are densely represented. In this perspective, neutral and full expressions are two prototypical categories characterized by high distinctiveness, while subtle facial expressions of anger might be characterized by a low distinctiveness and, as a consequence, might elicit smaller SPCN amplitudes and be associated with reduced accuracy in the change detection task when compared to full expressions of anger and neutral expressions.

Finally, we want to examine the lack of an impact of the mimicry manipulation on overt behavior. This discrepancy between the neural and behavioral levels of our investigation could originate from at least two possible sources. On the one hand, it is possible that the neural measure might be more sensitive to the mimicry manipulation than accuracy, at least in the context of the present change detection paradigm. According to this line of reasoning, the effect size of the mimicry effect on SPCN amplitude values was lower than the effect size of the emotion effect ($\eta p^2 = .161$ vs. $\eta p^2 = .267$), an observation that could suggest that accuracy captured only the greatest effect in terms of effect size. The literature offers several examples of this discrepancy between neural and behavioral findings (e.g., Heil, Rolke, and Pecchinenda, 2004; Luck, Vogel, and Shapiro, 1996), also in the context of the change detection task (e.g., Sessa et al., 2011, 2012). Moreover, the effects related to mimicry tend to be very small and usually require, in behavioral studies, rather large samples (with $N$ even larger than 100) to be observed (e.g., Wood et al., 2015). An alternative explanation of this incongruity could be that SPCN and accuracy provide estimates of two different aspects of the VWM functioning: while the SPCN can be considered a pure index of VWM representation, accuracy also reflects the processes of retrieval, deployment of attention on the test array stimuli and the comparison between the stored representation and the to-be-compared stimulus in the test array (i.e., *comparison process*; (Awh, Barton, and Vogel, 2007; Dell'Acqua et al., 2010; Hyun et al., 2009). Therefore, in light of these observations, we believe that the most relevant result of the present investigation, related to a modulation of the SPCN amplitude as a function of the mimicry manipulation, is entirely reliable.

In conclusion, the present investigation showed that VWM representations of facial expressions are sensitive to the observer's mimicry and, more specifically, that mimicry, when it can be used freely, is able, as predicted by the model of Wood and colleagues (2016), to enhance the clarity of these representations, which in neural terms translate into greater SPCN amplitude values than when mimicry is prevented or altered. These results, therefore, support this aspect of Wood and colleagues' model by providing direct evidence of this relationship between mimicry and VWM representations, and further clarify that the specific stage of VWM is involved in this feedback processing from simulation to visual analysis. Moreover, our findings represent a progress also with regard to the studies on VWM as they provide further knowledge on how this memory system operates and on what can be the sources of input information able to modulate its functioning. Specifically, our results suggest that VWM receives the feedback of sensorimotor regions during the processing of faces and facial expressions, including also emotional information. We therefore believe, overall, that the present results can represent a valuable progress for the definition of a simulation model for the recognition and understanding of others' emotions.

**Author contributions**

P.S. developed the study concept. All authors contributed to the study design. A.S.L. performed testing and data collection. P.S. and A.S.L. performed the data analysis and all the authors interpreted the data. P.S. drafted the manuscript. R.L. provided critical revision. All authors approved the final version of the manuscript for submission.

**Competing financial interests**

The authors declare no competing financial interests.

**Acknowledgments**

We would like to thank Prof. Lesley Fellows for kindly providing us with the stimuli used in the present study (see Vaidya, Jin and Fellows, 2014).

**Figure Captions**

Figure 1. Examples of the stimuli used in the change detection task, one for each three level of facial expression (neutral, intermediate, full) for 2 individual faces (one of a female and one of a male).

Figure 2. Timeline of each trial of the change detection task.

Figure 3. Mean proportion of correct responses in the change detection task for each facial expression condition (neutral, subtle, full).

Figure 4. Grand averages of the face-locked ERP waveforms time-locked to the presentation of the memory array as a function of the facial expression conditions (neutral, subtle, full) collapsed across the mimicry conditions (free vs. altered/blocked).

Figure 5. Grand averages of the face-locked ERP waveforms time-locked to the presentation of the memory array as a function of the mimicry conditions (free vs. altered/blocked) and collapsed across facial expression conditions (neutral, subtle, full) for medium-low EQ participants (panel A) and medium-high EQ participants (panel B) separately.